\documentclass[12pt]{article}
\def\bc{\begin{center}}
\def\ec{\end{center}}
\def\beq{\begin{equation}}
\def\eeq{\end{equation}}

\title{\bf Odderon with a running coupling constant}
\author{M.A.Braun  \\
S.Peterburg State University, Russia
}
\begin{document}
\maketitle

\begin{abstract}
The running coupling is introduced into the equation for
the odderon via the bootstrap relation. It is shown that the
previously found odderon state with a maximal intercept, which is
constructed from antisymmetric pomeron wave function, continues to
exist in the running coupling case. Its intercept is found to remain 
equal to unity independent of the  behaviour assumed for the
running coupling at low momenta.
\end{abstract}

\section{Introduction}
Lately a renewed interest has been shown towards introduction of a running
coupling into the BFKL dynamics.
Summation of contributions from  quark-
antiquark loops to the evolution of the gluon density has been used to
restore the full dependence of the coupling on the running scale in
the color dipole approach ~\cite{kovchegov1, kovchegov2, balitsky}.
It turned out that the obtained kernel for the linear BFKL equation
essentially
coincides with the one which we found many years ago by imposing the
bootstrap relation
necessary for the fulfilment of unitarity  ~\cite{braun1,braun2}.
In this paper we draw
attention to the fact that the bootstrap relation in fact allows to derive
also the
structure of the odderon in the perturbative QCD approach.
It is of a particular interest to know what happens to the odderon
state with a maximal intercept unity found in ~\cite{BLV} and expressed
via an antisymmetric pomeron state when a running coupling is introduced.
We demonstrate that since the bootstrap relation plays a decisive role
in construction of this solution to the odderon equation, introduction
of the running coupling via the bootstrap allows to preserve this
particular solution. 

As in our earlier papers, we have to stress from the start that introduction
of the running coupling into the BFKL formalism cannot be made rigorously
and uniquely.
The formalism admits transverse momenta of any magnitude, including very
small ones,
for which the concept of the gluon and its coupling looses any  sense.
The introduced
running coupling has to be artificially continued to small momentum
values,
where it is completely undetermined.
As a result the actual values for the pomeron and hence odderon
intercepts generally depend on a chosen behaviour of the running coupling
in the region of very low momenta.
However particular odderon states constructed from antisymmetric pomeron states 
continue to have their maximal intercept exactly equal to unity, as in the fixed coupling case studied in ~\cite{BLV}.

Note that we use the bootstrap equation only
in the lowest order of the coupling. So one can
hope for only establishing the leading order behaviour in the running
coupling. For the pomeron,
 in the next-to-
leading order, it has been explicitly  found
that the bootstrap method correctly  reproduces the part
of the kernel  responsible for the running of the coupling but not
the rest piece ~\cite{bravanew, fadin}. One expects the same
to be true also for the odderon.

We limit ourselves to the case of the odderon at rest
(with zero total momentum). Generalization to non-zero total momenta is
straightforward.

\section{The running coupling from the bootstrap}
As mentioned in the Introduction, our idea to introduce a running
coupling via the bootstrap was presented in ~\cite{braun1,braun2}. It
consists in expressing both the gluon trajectory and intergluon
interaction  in terms of a single function $\eta(q)$ of the gluon
momentum in the way which preserves the bootstrap relation.
Function $\eta(q)$ then can be chosen to conform to the
high-momentum behaviour of the gluon density with a running
coupling. Let the two-gluon equation be written as an
inhomogeneous Schroedinger equation
\beq
(H-E)\psi =F,
\label {scheq}
\eeq
with the Hamiltonian for the colour group $SU(N_{c})$
\beq
H=\frac{1}{2}N_{c}\Big(t(q_{1})+t(q_{2})\Big)+(T_{1}T_{2})V.
\label{ham2}
\eeq
Here
$(N_{c}/2)t(q)\equiv-\omega(q)$ is the kinetic energy of the gluon
given by its Regge trajectory with the minus sign.
We write $t(q)$ in the form
\beq
t(q)=\int \frac{d^2q_1}{(2\pi)^2}\frac{\eta(q)}
{\eta(q_1)\eta(q-q_1)},
\label{traj}
\eeq
where $\eta(q)$ is an
arbitrary function satisfying the condition 
\beq \eta(0)=0,
\label{eta0} 
\eeq 
which guarantees that the gluon trajectory
passes through zero at $q=0$ in accordance with the gluon
properties. The interaction term, apart from the product of the
gluon colour vectors $T_{i}^{a},\ i=1,2,\ a=1,...N_{c}$, involves
the interaction kernel $V$ which we also express via the same
function $\eta(q)$
\beq
V(q_1,q_2|q'_1q'_2)=\frac{\eta(q_1+q_2)}{\eta(q'_1)\eta(q'_2)}
-\Big(\frac{\eta(q_1)}{\eta(q'_1)}+
\frac{\eta(q_2)}{\eta(q'_2)}\Big)\frac{1}{\eta(q_1-q'_1)}.
\label{int}
\eeq
It conserves the momentum: $q_1+q_2=q'_1+q'_2$.
Comparing (\ref{traj}) with (\ref{int}) one finds the bootstrap
relation, satisfied for arbitrary $\eta(q)$: 
\beq
\int\frac{d^{2}q_{1}}{(2\pi)^{2}}V(q_1,q_{2}|q'_1,q'_2)=
t(q_{1})+t(q_{2})-t(q_{1}+q_{2}). 
\label{boots} 
\eeq

From the high-momentum behaviour of the gluon distribution with a running
coupling one finds
\beq
\eta(q)=\frac{1}{2\pi}bq^2\ln\frac{q^2}{\Lambda^2},\ \ q^2>>\Lambda^2,
\label{asym}
\eeq
where $\Lambda$ is the standard QCD parameter and
\beq
b=\frac{1}{12}(11 N_c-2N_f).
\label{bval}
\eeq
The asymptotic (\ref{asym}) and condition (\ref{eta0}) are the only
properties of $\eta(q)$ which follow from the theoretical reasoning.
A concrete form of $\eta(q)$ interpolating between (\ref{eta0}) and
(\ref{asym}) may be chosen differently. One hopes that
physical results will not too strongly depend on the choice.

The fixed coupling case corresponds to the choice
\beq
\eta(q)=\frac{2\pi}{g^2}q^2.
\label{etafix}
\eeq
Then one finds the standard BFKL expressions for the trajectory $t(q)$
and interaction $V(q_1,q_2|q'_1,q'_2)$.

It is customary to discuss a homogeneous equation (\ref{scheq})
(with $F=0$) to
seek for the ground state energy, which determines the rightmost
singularity in the complex angular momentum $j$ related to the energy by
\beq j=1-E \eeq
For our purpose it is, however, essential to conserve the inhomogeneous
term $F(q_{1},q_{2})$, which represents the two-gluon-particle vertex.

Consider the vector colour channel with $T_1T_2=-N_c/2$ and, most
important, assume that the inhomogeneous term depends only on the total
momentum of the two gluons: $F=F(q_1+q_2)$. Then using the bootstrap
relation (\ref{boots}) one easily finds the solution to Eq. (\ref{scheq}):
\beq
\psi(q_1,q_2)=\psi(q_1+q_2)=\frac{F(q_1+q_2)}{\omega(q_1+q_2)-E}.
\eeq
It means that the two gluons 1 and 2 have fused into a single one with
the momentum $q_1+q_2$. This phenomenon is true for aribitrary $\eta(q)$.

\section{The odderon}
\subsection{The bootstrap for three gluons}

Noe we generalize the bootstrap for two gluons to the
case of three gluons, relevant for the odderon.
For three gluons Eq. (\ref{scheq}) holds with the
Hamiltonian which is a sum of kinetic energies and pair interactions:
\beq
H=\frac{1}{2}N_{c}\sum_{i=1}^{3}t(q_{i})+\sum_{i<k}^{3}(T_{i}T_{k})V_{ik}.
\eeq
Here $V_{ik}$ is the interaction of gluons $i$ and $k$ with the kernel
$V(q_i,q_k|q'_i,q'_k)$. We do not impose any restrictions on the
total colour $T=\sum_{i=1}^{3}T_{i}$. (However only for $T=0$ the
Hamiltonian is infrared stable).

Assume now that $T_{1}T_{2}=-N_{c}/2$, i.e. the gluons 1 and 2 form a
colour vector. Then we demonstrate that for a certain specific choice of the
inhomogeneous term $F$  Eq. (\ref{scheq}) for $3$ gluons reduces to that
for two
gluons, the gluons 1 and 2 fused into a single gluon which carries their
total momentum and colour. In other words, as in the two-gluon case, a
pair
of gluons in the adjoint representation is equivalent to a single gluon.

Of course, the specific form of the inhomogeneous term is a decisive
instrument for bootstrapping the two gluons 1 and 2. For three gluons we
choose
\beq
F_{3}(q_{1},q_{2},q_{3})=\int \frac{d^{2}q'_{3}}{(2\pi)^{2}}
\hat{W}(q_{1},q_{2},q_{3}|q'_{1},q'_{3})\psi_{2}
(q'_{1},q'_{3})+F_{2}(q_{1}+q_{2},q_{3}),
\label{inho}
\eeq
with $q'_{1}+q'_{3}=q_{1}+q_{2}+q_{3}$. In Eq. (\ref{inho})
$\psi_{2}$ is a
solution of the Schroedinger equation (\ref{scheq})
with the inhomogeneous term $F_2$. Gluon 1 in it substitutes
the fused initial gluons 1 and 2.
The kernel $\hat{W}$ is an operator acting also on colour indeces of
$\psi_2$. It is convenient to retain a pair of colour indeces for
gluon 1 in $\psi_2$ inherited from the initial gluons 1 and 2 by
means of a projector onto the adjoint representation in the colour
$T_{1}+T_{2}$. This allows to consider $\hat{W}$ as an operator acting
in the colour space of three gluons (not two). Then it has the form
\beq
\hat{W}(q_{1},q_{2},q_{3}|q'_{1},q'_{i})=
(T_{1}T_{3})W(q_{2},q_{1},q_{3}|q'_{1},q'_{3})+
(T_{2}T_{3})W(q_{1},q_{2},q_{3}|q'_{1},q'_{3}),
\label{opw}
\eeq
where the momentum space kernel $W$ is a difference between two kernels
(\ref{int}), expressed via function $\eta(q)$:
\beq
W(q_{2},q_{1},q_{3}|q'_{1},q'_{3})=
V(q_{1},q_{3}|q'_{1}-q_{2},q'_{3})
-V(q_{1}+q_{2},q_{3}|q'_{1},q'_{i}).
\label{opw1}
\eeq
The explicit form of $W$ is irrelevant for our purpose. In terms
of function $\eta(q)$ it can be found in ~\cite{braunew}. It possesses
a symmetry property, which is important for us:
\beq
W(q_{1},q_{2},q_{3}|q'_{1},q'_{3})=
W(q_{3},q_{2},q_{1}|q'_{3},q'_{1}).
\label{symw}
\eeq
Note that in the fixed coupling case this kernel reduces to
the Bartels kernel $K_{2\to 3}$ which describes transition from two
BFKL pomerons to three ~\cite{bartels}.

We are going to show that, with the inhomogeneous term $F_{3}$
given by Eq. (\ref{inho})
and $T_{1}T_{2}=-N_{c}/2$, the Schroedinger equation (\ref{scheq})
for three gluons is solved by
\beq
\psi_{3}(q_{1},q_{2},q_{3})
=\psi_{2}(q_{1}+q_{2},q_{3}).
\label{sol}
\eeq
Indeed putting (\ref{sol}) into the equation we find that the
interaction term $(T_{1}T_{2})V_{12}$, according to (\ref{boots}),
substitutes the sum of kinetic
terms $t(q_{1})+t(q_{2})$ for the gluons 1 and 2 by $t(q_{1}+q_{2})$,
which is precisely the kinetic term for the function
$\psi_{2} (q_{1}+q_{2})$. The interaction of gluon 1 with the third one
takes the form
\[
(T_{1}T_{3})\int \frac{d^{2}q'_3}{(2\pi)^{2}}V(q_{1},q_{3}|q'_{1},q'_{3})
\psi_{2}(q'_{1}+q_{2},q'_{3})=\]
\beq (T_{1}T_{3})\int\frac
{d^{2}q'_{3}}{(2\pi)^{2}}V(q_{1},q_{3}|q'_{1}-q_{2},q'_{3})
\psi_{2}(q'_{1},q'_{3}).
\label{term13}
\eeq
The momentum is conserved during the interaction, so that on
the lefthand side $q_{1}+q_{3}=q'_{1}+q'_{3}$ and on the righthand side
$q_{1}+q_{2}+q_{3}=q'_{1}+q'_{3}$.  The term (\ref{term13})
is cancelled by the
identical contribution with the opposite sign coming from the
inhomogeneous term $F_{3}$ (the first term in (\ref{opw1}) in the part of
$\hat{W}$ proportional to $T_{1}T_{3}$). Instead of it from the
inhomogeneous term comes the contribution (the second term in (\ref{opw1}))
\beq
(T_{1}T_{3})
\int \frac{d^{2}q'_{3}}{(2\pi)^{2}}V(q_{1}+q_{2},q_{3}|q'_{1},q'_{3})
\psi_{2}(q'_{1},q'_{3}).
\label{int13}
\eeq
In the same manner the interaction of gluon 2 with the third one
\[
(T_{2}T_{3})\int \frac{d^{2}q'_{3}}{(2\pi)^{2}}V(q_{2},q_{3}|q'_{2},q'_{3})
\psi_{2}(q_{1}+q'_{2},q'_{3})=\]
\beq (T_{2}T_{i})\int\frac
{d^{2}q'_{3}}{(2\pi)^{2}}V(q_{2},q_{3},q'_{2}-q_{1},q'_{3})
\psi_{2}(q'_{2},q'_{3})
\eeq
is transformed by the term in $F_{3}$ proportional to $T_{2}T_{3}$ into
\beq
(T_{2}T_{3})
\int \frac{d^{2}q'_{3}}{(2\pi)^{2}}V(q_{1}+q_{2},q_{3}|q'_{1},q'_{3})
\psi_{2}(q'_{1},q'_{3}).
\label{term23}
\eeq
The two terms (\ref{term13}) and (\ref{term23}) sum into
\beq
(T_{1}+T_{2},T_{3})
\int \frac{d^{2}q'_{3}}{(2\pi)^{2}}V(q_{1}+q_{2},q_{3}|q'_{1},q'_{3})
\psi_{2}(q'_{1},q'_{3}),
\eeq
which is precisely the correct form for the interaction of the gluon
with the total colour $T_{1}+T_{2}$ and momentum $q_{1}+q_{2}$
in which the initial gluons 1 and 2 have fused.
As a result we obtain
the Schroedinger equation (1) for the function $\psi_{2}$ describing
two gluons with the inhomogeneous term $F_{2}$.

The inhomogeneous term of the form (\ref{inho}) actually appears in
the theory of three reggeized gluons, provided one takes into account the
possibility for transitions from 2 to 3 gluons. In the case of
the fixed coupling constant this fact was discovered
in the course of the perturbative analysis of the system of
4 reggeized gluons in the triple Regge kinematical region
~\cite{bartels1, BW}.
Introducing a running coupling via the bootstrap allows to preserve
this result for a general choice of function $\eta(q)$.

We stress that this result is valid for any value of the total colour
of the three-gluon system and any choice of the symmetry of the
spatial wave function. In the particular case of the colorless three-gluon
system with a $d$-structure of the color wave function (the odderon state)
the bootstrap allows to find a specific solution to the odderon equation,
expressed via an antisymmetric pomeron state.

\subsection{A solution of the odderon equation}
In the case of the odderon the total colour is zero, so that any pair of
gluons is in the colour vector state and $T_iT_k=-N_c/2$ for $i<k=1,2,3$.
The total color wave function is $d_{a_1a_2a_3}$ where $a_i$ is the
color of the $i$-th reggeized gluon. Separating this color factor one
finds the  equation for the odderon momentum wave function
\beq
(H_{od}-E)\psi_{od}=0,
\label{odeq}
\eeq
with a Hamiltonian
\beq
H^{od}=\frac{1}{2}N_{c}(\sum_{i=1}^{3}t(q_{i})-\sum_{i<k}^{3}V_{ik}).
\label{hamod}
\eeq

Now we use the bootstrap results obtained above. Consider the odderon
equation with the inhomogeneous term (\ref{inho}) in which $T_1T_3=T_1T_2=
-N_c/2$, the colour wave function $d_{a_1a_2a_3}$ and  $F_2=0$.
Our result tells us
that this equation is solved by the pomeron momentum wave function
\beq
\psi^{(12)}=\psi_2(q_1+q_2,q_3)
\eeq
which satisfies the homogeneous Schroedinger equation (\ref{scheq})
with a Hamiltonian   (\ref{ham2}) in which $T_1T_2=-N_c$ (the BFKL equation
in the case of a fixed coupling). Explicitly we find that
in the momentum space
\beq
(H_{od}-E)\psi^{(12)}=F_3^{(12)},
\label{odeq12}
\eeq
where
\beq
F_{3}^{(12)}(1,2,3)=-\frac{1}{2}N_c\int \frac{d^{2}q'_{3}}{(2\pi)^{2}}
\Big(W(2,1,3|1',3')+W(1,2,3|1',3')\Big)\psi_{2}(1',3').
\label{inho12}
\eeq
Here and in the following for brevity we  denote gluon momenta just
by numbers: $1\equiv q_1$, $1'\equiv q'_1$ etc.

Next we cyclically permute the gluons 1,2,3 to obtain two more relations
\beq
(H_{od}-E)\psi^{(23)}=F_3^{(23)}
\label{odeq23}
\eeq
and
\beq
(H_{od}-E)\psi^{(31)}=F_3^{(31)},
\label{odeq31}
\eeq
where for instance
\beq
\psi^{(23)}=\psi_2(q_2+q_3,q_1)
\eeq
and
\beq
F_{3}^{(23)}(1,2,3)=-\frac{1}{2}N_c\int \frac{d^{2}q'_{3}}{(2\pi)^{2}}
\Big(W(3,2,1|1',3')+W(2,3,1|1',3')\Big)\psi_{2}(1',3')
\label{inho23}
\eeq
(the integration momenta $q'_3$ and $q'_1=q_1+q_2+q_3-q'_3$ do not change
under permutations of the external momenta).
We add all the three Schroedinger equations obtained in this manner
together  to obtain
\beq
(H_{od}-E)\Big(\psi^{(12)}+\psi^{(23}+\psi^{(31)}\Big)=
F_3^{(12)}+F_3^{(23)}+F_3^{(31)} \equiv F_3^{tot}.
\label{odeqtot}
\eeq
The total inhomogeneous term has a structure
\beq
F_{3}^{tot}(1,2,3)=-\frac{N_c}{2}\int\frac {d^{2}q'_{3}}{(2\pi)^{2}}
U(1,2,3|1',3')\psi_{2},
(1',3')
\label{inhotot}
\eeq
where
\[
U(1,2,3|1',3')=
W(2,1,3|1',3')+W(1,2,3|1',3')+W(3,2,1|1',3')\]\beq
+W(2,3,1|1',3')+
W(1,3,2|1',3')+W(3,1,2|1',3').
\eeq
Using property (\ref{symw}) one easily finds that $U$ is symmetric in the
last pair of arguments
\beq
U(1,2,3|1',3')=U(1,2,3|3',1').
\eeq
It follows that if function $\psi_2(q_1,q_2)$ is antisymmetric in its
variables, the total inhomogeneous term  (\ref{inhotot}) in Eq.
(\ref{odeqtot}) vanishes and the sum
\beq
\psi_{od}=\psi^{(12)}+\psi^{(23}+\psi^{(31)}
\label{odsol}
\eeq
is a solution of the odderon equation
\beq
(H_{od}-E)\psi_{od}=0.
\label{eqod}
\eeq
Thus from any antisymmetric soloution of the pomeron equation one gets
a corresponding solution of the odderon equation.

For a fixed coupling this result was found in ~\cite{BLV}. Our derivation
generalizes it to the case of a running coupling, provided it is
introduced in the manner which preserves the bootstrap relations.

\section{The pomeron and odderon intercepts}
With the running coupling constant the intercept of the pomeron states
generally begins to depend on the way in which the constant is chosen
to behave at small values of momenta (where in fact it is not defined physically).
We recall that our function $\eta(q)$ is fixed only by its asymptotic behaviour
at large values of $q$ (\ref{asym}) and its value at $q=0$ (\ref{eta0}).
A possible simple form of $\eta(q)$ satisfying these conditions and
no more singular at $q=0$ than in the fixed coupling case is
\beq
\eta(q)=\frac{1}{2\pi}bq^2\ln\Big(\frac{q^2}{\Lambda^2}+\xi^2\Big),
\label{etaq}
\eeq
with two parameters, the standard QCD parameter $\Lambda$ and $\xi>1$,
constrained by the experimental value of the coupling constant $\alpha_s(q^2)$ at
some chosen $q^2$ value. This form implies that the running coupling $\alpha_s(q)$
steadily grows towards small values of $q$ and freezes at $q\sim q_0\sim\xi\Lambda$
to have its value at $q=0$ 
\beq
\alpha_s(0)=\frac{\pi}{2b\ln\xi}.
\eeq
Choosing $m_w$ as the experimental scale and taking  $\alpha_s(m_w^2)=0.126$
(to have $\Lambda\simeq 0.2$  GeV/c at small values of $\xi$, in the one-loop approximation) we find that the interval of possible values of $\xi$ is
$1\leq\xi\leq \xi_{max}=1.57\cdot 10^5$. As $\xi$ rises, so do $\Lambda$ and $q_0$, whereas
$\alpha_s(0)$ and the intercept $\Delta$ steadily diminish towards their limiting
values at $\xi_{max}$, 0.126 and 0.330 respectively. Values of all these quantities at different $\xi$ are presented in the Table.
\vspace*{0.5 cm}
\begin{center}
{\bf Table. Pomeron intercepts}
\vspace*{0.5 cm}

\begin{tabular}{|r|r|r|r|r|}
\hline
$\xi$ & $\Lambda$ (GeV/c)&$\alpha_s(0)$&$q_0$ (GeV/c)&$ \Delta$\\\hline
2.0 &  0.197 & 2.18& 0.278 & 5.69\\
2.0$\cdot 10^2$ & 0.197 & 0.284 & 2.78 & 0.743 \\
2.0$\cdot 10^3$ & 0.198 & 0.198 & 8.84 & 0.516 \\
2.0$\cdot 10^4$ & 0.210 & 0.152 & 29.7 & 0.395 \\
5.0$\cdot 10^4$ & 0.237 & 0.139 & 53.2 & 0.361 \\
9.0$\cdot 10^4$ & 0.300 & 0.132 & 90.0 & 0.343 \\
1.5$\cdot 10^5$ & 0.893 & 0.127 & 346  & 0.330 \\\hline
\end{tabular}
\end{center}
\vspace*{0.5 cm}

In this respect it is worth noting that introduction of the running
coupling does not eliminate the arbitrariness in the value chosen for the
 coupling in the fixed coupling case,
just trading this arbitrariness for the one in the choice of, say,
$\alpha_s(0)$.

However it is remarkable that introduction of the running coupling
does not change the maximal intercept of the odderon, which, with unity
subtracted, remains equal to zero.
In the fixed coupling case the maximal intercept of antisymmetric
pomeron states is unity ($E=0$). Correspondingly the odderon built
from these states according to Eq. (\ref{odsol}) has its maximal intercept
equal to unity. It is trivial to show that introduction of a running coupling
does not change this result.

Indeed  consider the Schroedinger equation (\ref{scheq}) for the forward
pomeron: $q_1+q_2=0 $, $T_1T_2=-N_c$. Divided by $N_c$, in terms of
function $\eta(q)$ this equation for the semi-amputated function
\beq
\phi(q)=\eta(q)\psi_2(q)
\eeq
has the form
\beq
\phi(q)\int \frac{d^2q_1}{(2\pi)^2}\frac{\eta(q)}
{\eta(q_1)\eta(q-q_1)}-
2\int\frac{d^2q_1}{(2\pi)^2}\frac{1}{\eta(q-q_1)}\phi(q_1)=E\phi(q).
\label{scheq1}
\eeq
This equation has an obvious antisymmetric solution
\beq
\phi(q)=\frac{{\bf q}}{\eta(q)}
\label{phi}
\eeq
corresponding to the eigenvalue $E=0$.
Indeed with (\ref{phi}) the left-hand side of (\ref{scheq1}) acquires the form
\beq
{\bf q}\int \frac{d^2q_1}{(2\pi)^2}\frac{1}
{\eta(q_1)\eta(q-q_1)}-
2\int\frac{d^2q_1}{(2\pi)^2}\frac{{\bf q}_1}{\eta(q_1)\eta(q-q_1)}.
\label{scheq2}
\eeq
In the second term,  the change of variables $q_1\to q-q_1$
transforms the numerator into ${\bf q}-{\bf q_1}$.
The half of the sum of these equal expressions coincides with the first
term in (\ref{scheq2}) so that the left-hand side vanishes. Thus
(\ref{phi}) is indeed a solution of Eq. (\ref{scheq1}) with a zero
eigenvalue.

\section{Conclusions}
Introduction of the running coupling via the bootstrap relation
advocated for the pomeron a long time ago in ~\cite{braun1,braun2}
allows to construct in a straightforward manner also an equation 
for the odderon with a running coupling. Its form admits  solutions
expressed via antisymmetric pomeron states, similar to the ones
found in  the fixed
coupling case. The maximal value of the intercept for such
solutions remains equal to unity as in the fixed coupling case.

\section{Acknowledgements}
This work has been supported by grants RNP 2.1.1.1112
and RFFI 06-02-16115a of Russia.

\end{document}